\documentclass{article}

\usepackage{amsmath}
\usepackage{tikz}
\usetikzlibrary{arrows}
\usetikzlibrary{shapes,snakes}

\begin{document}

\title{Scalable and automated Evaluation of Blue Team cyber posture in Cyber Ranges}

\author{Federica Bianchi, Enrico Bassetti, Angelo Spognardi}
\date{}

\maketitle

\begin{abstract}
Cyber ranges are virtual training ranges that have emerged as indispensable environments for conducting secure exercises and simulating real or hypothetical scenarios. These complex computational infrastructures enable the simulation of attacks, facilitating the evaluation of defense tools and methodologies and developing novel countermeasures against threats. One of the main challenges of cyber range scalability is the exercise evaluation that often requires the manual intervention of human operators, the White team. This paper proposes a novel approach that uses Blue and Red team reports and well-known databases to automate the evaluation and assessment of the exercise outcomes, overcoming the limitations of existing assessment models. Our proposal encompasses evaluating various aspects and metrics, explicitly emphasizing Blue Teams' actions and strategies and allowing the automated generation of their cyber posture.

\end{abstract}

\section{Introduction}

In the face of escalating cyber threats, organizations must continually fortify their defenses against malicious actors. Cyber range exercises have emerged as vital training grounds, enabling security teams to simulate realistic attack scenarios and assess their preparedness. Typically involving Blue Teams defending infrastructure and Red Teams launching simulated attacks, these exercises offer hands-on experience. However, evaluating exercise results has been challenging, relying on a combination of service metrics and manual grading by human experts. This approach is time-consuming, error-prone, and limits prompt feedback on Blue Team responses. Also, the grading phase often receives little attention, hindering the potential for comprehensive improvement of defensive strategies~\cite{yamin2020cyber,yamin2022}. As cybersecurity becomes increasingly paramount, there is a pressing need for a robust evaluation metric, automated processes, and objective insights into Blue Team performance.

To address these limitations, this paper presents a novel approach that leverages reports and well-known cybersecurity databases to automate the evaluation and assessment of cyber range exercise results, explicitly focusing on the Blue Team's performance. Our proposed framework exploits the power of automation and addresses the limitations of manual evaluation methods to offer an efficient and objective means of assessing various aspects and metrics related to the exercise.

The primary contribution of this research lies in developing a comprehensive and scalable approach that automates the evaluation process. By producing a tree-based representation of attack and defense reports, we propose to evaluate a cyber exercise leveraging the comparison between the models of the different Red and Blue teams' actions. Our approach allows the automated comparison and evaluation of multiple Blue teams in parallel, helping the White team in the exercise evaluation and enabling organizations to extract valuable insights from the results promptly.
The contribution provided concerns, among other things, the possibility of evaluating various aspects and metrics related to the Blue Team, particularly those related to the correctness and accuracy of the response to Red Team attacks.

This paper is structured as follows: Section~\ref{sec:background} provides a background of the current evaluation methods and their limitations; Section~\ref{sec:automatic-evaluation} provides the general idea for our proposal; Sections~\ref{sec:team-reports}, \ref{sec:reports-to-adtree} and~\ref{sec:evaluation} describe in details the core of our proposal, i.e., how to define the cyber range reports for the teams and how to process them. Finally, Section~\ref{sec:conclusions} concludes the paper and discusses ideas for future research.

\section{Background and related works}\label{sec:background}
The current evaluation methods in cyber ranges vary in their approaches, with a growing emphasis on automating scoring to reduce manual efforts. Typically, cyber ranges display exercise progress through dashboards, and the evaluation is primarily based on the number of successfully completed challenges. Some platforms, such as I-tee~\cite{willems2012online}, Hack The Box (HTB)~\cite{htb}, iCTF~\cite{iCTF}, Kypo~\cite{vceleda2015kypo}, and Locked Shields~\cite{lockedshields} use scoring systems based on goal completion and additional metrics like service availability, integrity or confidentiality.

However, the limitations of these methods are evident. Automatic scoring engines rely on basic calculations tied to points obtained achieving specific objectives or on metrics, neglecting the detailed performance and strategies employed by participants. Such evaluations often lack a comprehensive assessment of participants' capabilities.

Andreolini et al.~\cite{andreolini2020framework} proposed a scoring system that evaluates participants' actions by comparing them against ``ideal'' actions defined by instructors. This involves constructing reference and trainee graphs representing ideal and participant actions. While a step forward, this approach has its own set of limitations.

Firstly, it may not fully capture the strategies and defenses of the Blue team, mainly since they depend on the red team's attacks, which can be decided ongoing and thus cannot be modeled \textit{a priori} from the graphs entered by the instructors. Secondly, the evaluation process is not fully automatic, requiring continuous manual work by instructors to define reference graphs and adapt them to evolving exercises.

The subsequent chapters will describe an alternative approach that aims to automatically evaluate the Blue team's results with minimal manual contribution by the white team.

\section{Automatic evaluation of exercises}\label{sec:automatic-evaluation}

We propose a new approach for automatically evaluating cyber-range exercises, focusing on the Blue Team results. This new approach comprises well-defined templates for Red and Blue Teams reports, a set of procedures to derive the final score, and a visualization named \textit{Cyber Posture}.

The pipeline for the automatic evaluation can be formalized as follows: (1) collection of reports from Blue and Red Teams; (2) definition of Reference/Response Graphs from reports; (3) automatic evaluation of multiple intermediate scores from defined graphs; (4) computation of the final score and the Cyber Posture.

We start by defining the structure for the reports. Then, we explain how to automatically build Reference and Response Graphs from team reports when the exercise ends: both are necessary to the score definition. Finally, we explain how to compute scores and the Cyber Posture from graphs data.

\section{Team reports}\label{sec:team-reports}
Most modern cyber ranges are equipped with a \textit{reporting system}, which generates reports during the exercise. Reports are produced by the Red Team when they attack and are matched with reports from the Blue Team describing the detected attack. 
Since one task of the White Team is to evaluate the Blue Teams based on these reports, our goal is to lighten the load of the White Team for the evaluation by making these phases automatic.

The structures we propose for Blue and Red Team reports are based on the components of the MITRE ATT\&CK~\cite{strom2018mitre} Matrix (accessed via STIX~\cite{barnum2012standardizing}), a database containing knowledge collected by the security community about tactics, techniques, and procedures used by attackers, together with the corresponding possible mitigations and detections. Reports structures prioritize simplicity, ease of filling, and unobtrusiveness, especially for the Blue Team, enhancing and efficient and effective report writing and the accuracy of automated analysis. 

The Red Team report template includes fields such as attack objective, techniques, sub-techniques, target of the attack, start time, and outcome. It could also arbitrarily contain \textit{Desirable Mitigation} and the \textit{Desirable Detection}, which are filled by the White Team either at the beginning of the exercise, or at the end of the simulation, before the automatic evaluation phase is initiated.

Additionally, the White Team can also choose to assign weights in $[0,1]$ to specific fields in the report- \textit{Tactic}, \textit{Techniques}, \textit{Sub-Techniques}, \textit{Desirable Mitigations}, and \textit{Desirable Detection}. These weights impact the Blue Team evaluation by assigning priorities to specific aspects that they may want to better assess. If unassigned, automatic weights will be applied during evaluation (Section~\ref{sec:evaluation}).

The \textit{Desirable Detection} and \textit{Desirable Mitigations} optional fields only gain significance if the White Team assigns weights to the abovementioned fields.
In this way, a Blue Team achieves an higher score by precisely executing one of those mitigations and detections. Anyways, even if it used a mitigation or detection technique not listed by the White Team but still present in the ATT\&CK for that attack, it would still take a higher score than if it had not performed proper mitigation or detection.

The Blue Team report template consists of presumed tactics, techniques, sub-techniques, applied mitigations, detection types, target attacked, and detection start time. 
\section{From reports to ADTrees}\label{sec:reports-to-adtree}

Once the teams fill the reports, the cyber range scoring system processes them to produce two graphs, \textit{Reference Graph} and \textit{Response Graph}, for each report. These two graphs are used in the scoring phase to calculate the total score assigned to each Blue Team. We named the structure of these graphs \textit{ReportADTree}, as they are constructed as a variant of ADTree~\cite{kordy2014attack,bagnato,schneier1999features}, a directed tree modeling attack-defense scenarios using two types of nodes: \textit{attack} and \textit{defense} nodes. 

The attack nodes consist of tactics, techniques, and sub-techniques nodes, as defined in the MITRE ATT\&CK. In ReportADTrees, the initial vertex is always the tactic, followed by techniques, which can branch into mitigations, detections, and sub-techniques. Sub-techniques, in turn, may have their own mitigations and detections as child nodes.

We define \textit{Reference Graph} as a reference for the evaluation of attack response and detection by the exercise participants. The graph is assembled using the attack information from the Red Team report and the ATT\&CK database, following the ReportADTree structure. Each ReportADTree node has the weight assigned by the White Team in the Red Team report or automatically generated ones if not present.

We also define the \textit{Response Graph} that evaluates the Blue Team's response to Red Team attacks: it is compared to the corresponding reference graph to assign a score to the Blue Team. The Response Graph is constructed using the defense information extracted from the fields of a Blue Team report, with a process similar to that described for the Reference Graph. However, only the \textit{Mitigations} and \textit{Detection} defined by the Blue Team report are added as child nodes of the Techniques or Sub-techniques they refer to.

\section{Evaluation}\label{sec:evaluation}

The automatic evaluation phase starts once the Reference and Response graphs have been constructed. We propose to evaluate different factors to capture the variety of aspects of an exercise. We define multiple intermediate scores and an aggregated final score for each Blue Team evaluation. Factors include attack management, attack strategy comprehension, knowledge of (sub-)techniques, accuracy in identifying techniques, responsiveness, and metrics such as availability and integrity. The evaluation aims to measure the Blue Team's ability to respond to Red Team attacks, understand attack strategies, and accurately identify and mitigate techniques, ultimately contributing to an aggregated final score.

The evaluation starts with an initial phase, which aims to modify the reference and the response graphs to enable a direct evaluation based solely on response graph nodes and their associated weights. The White Team can specify the weights manually; they are automatically generated if missing.

Once all reference graph weights are set, the two graphs are compared using a breadth-first search to assign weights to response graph nodes that the Blue Team has correctly individuated. In contrast, nodes without matches in the reference graph are removed. For techniques and sub-techniques, even if there is not an exact match with the response graph, we can still estimate the accuracy of the guesses of the Blue Team. The idea is to compute the distance between the correct Technique and the one guessed by the Blue Team, using CAPEC~\cite{capec} from MITRE, a public database of known attack patterns, via existing CAPEC-ATT\&CK mappings~\cite{attackcapec}.

The score calculation involves defining multiple intermediate scores to evaluate factors mentioned earlier, with the option to associate weights with each intermediate score based on exercise objectives. Once all these intermediate scores are computed, they are averaged into a unique \textit{Final Score}.

The first intermediate score is the \textit{Comprehension Score}, an indicator that measures the \textit{knowledge of (sub-)techniques} and the \textit{attack strategy comprehension}. It assesses the Blue Team's precision in identifying the techniques and sub-techniques used by the Red Team in the attack and the Blue Team's comprehension of the overall attack strategy and tactic. The second one is the \textit{Defense Score}; it represents the \textit{attack management} factor, which is the Blue Team's ability to understand how to respond to Red Team attacks. The third is the \textit{Implementation Score}, which assesses whether the Blue Team implemented the identified mitigation. While the Comprehension Score evaluates the team's ability to identify the correct mitigations for techniques or sub-techniques, the Implementation Score focuses explicitly on the actual execution of those mitigations. Finally, the last score is the \textit{Responsiveness Score}, which represents the interval between the attack beginning time reported in the Red Team report and the beginning time reported in the Blue Team report.

From the information gathered, it is possible to define an overall picture of the capabilities that each Blue Team developed during the exercise, in addition to individual scores. This aspect is often referred to as \textit{Cyber Posture}~\cite{ecso-pdf}. The measures reflect the Blue Team's overall defensive ability, offering a comprehensive view of the exercise outcomes and individual evaluation aspects.

\section{Conclusions and future work}\label{sec:conclusions}

We presented a novel approach to automate the evaluation and assessment of cyber range exercises, explicitly focusing on Blue Teams. The proposed framework leverages custom reports, graphs, and well-known databases, such as MITRE ATT\&CK and CAPEC, to provide a comprehensive and scalable solution that addresses the limitations of traditional manual evaluation methods. The automation of the evaluation phase addresses the challenges of time-consuming and error-prone manual evaluations.


As future work, we are designing a fully working scoring platform for cyber ranges, integrating a GUI for receiving and evaluating the reports. Moreover, we plan to refine and expand the evaluation metrics used in the automated framework, identifying additional indicators and benchmarks that can provide a more comprehensive assessment of the Blue Team's performance. Finally, we plan to integrate machine learning and artificial intelligence techniques to enhance the automation process by enabling intelligent analysis and interpretation of the evaluation results.

\section{Acknowledgments}

This work was partially supported by project ``Prebunking: predicting and mitigating coordinated inauthentic behaviors in social media'' project, funded by Sapienza University of Rome.

\bibliographystyle{abbrv}
\bibliography{bibliography} 

\end{document}